\documentclass[useAMS,usenatbib,fleqn]{mn2e}
\usepackage{graphics,graphicx,times}
\usepackage{amsmath}
\usepackage{amssymb}
\usepackage{aas_macros}
\usepackage[usenames,dvipsnames]{xcolor}
\newcommand{\kms}{${\rm km\ s}^{-1}$\ }
\newcommand{\mpcc}{${\rm m_p\ cc}^{-1}$\ }

\def\Osix{O\hskip0.25mm{\sc vi}\ }

\newcommand{\rfree}{r_{\scriptscriptstyle \rm f}}
\newcommand{\tfree}{t_{\scriptscriptstyle \rm f}}
\newcommand{\epsa}{\epsilon}

\usepackage[para]{footmisc}

\title[Galactic wind interaction with halo gas]{Interaction of galactic wind with halo gas and the  origin of multiphase extraplanar material} 

\author[Sharma et al.]
{
Mahavir Sharma$^{1}$\thanks{mahavir@rri.res.in}, Biman B. Nath$^{1}$\thanks{biman@rri.res.in}, 
Indranil Chattopadhyay$^{2}$ and Yuri Shchekinov$^{3}$\thanks{yus@sfedu.ru} \\
$^{1}$ Raman Research Institute, Sadashivanagar, Bangalore 560080, India\\
$^{2}$ Aryabhatta Research Institute of Observational Sciences, Manora Peak, Nainital 263129, India\\
$^{3}$ Department of Physics, Southern Federal University, Rostov on Don 344090, Russia
}

\begin{document}

\date{Submitted ---------- ; Accepted ----------; In original form ----------}


\maketitle

\begin{abstract}
{
We study the interaction of galactic wind with hot halo gas using hydrodynamical simulations. We find that the outcome of this interaction depends crucially on the wind injection density and velocity. Various phases of the extraplanar media such as high velocity clouds (HVCs), outflowing clouds, and \Osix regions can originate in the interaction zones of wind with the  halo gas, depending on the injection velocity and density.   
In our simulations the size of the HVCs is of the order of $100$ pc. The total mass contained in the clouds is $10^5\hbox{--}10^7$ M$_\odot$ and they have a  normal distribution of velocities in the galactic standard of rest frame, similar to HVCs. For high injection density and velocity,  a significant number of clouds move outwards and resemble the case of cold neutral outflows. 
 Furthermore a $10^5\hbox{--}10^6$ K phase is formed in our simulations which has a column density $\sim 10^{18}$ cm$^{-2}$, and resembles the observed \Osix regions. 
The injection velocity and density are linked with the mass loading factor of the outflow, efficiency of energy injection due to supernovae and the SFR. 
Comparison of the predicted morphology of extraplanar gas with observations can serve as  a useful diagnostic for constraining feedback efficiency of outflows. }
 
\end{abstract}
\begin{keywords}
galaxies : evolution --  galaxies : haloes -- galaxies : starburst -- X-rays : galaxies   
\end{keywords}
\section{Introduction}
Disc galaxies contain a substantial amount of baryonic matter outside the disc, and this
extra-planar material has become an important problem to study for a number of reasons.
Theoretically, since the pioneering papers by \citet{Spitzer1956} and \citet{Pikelner1958} that invoked the
presence of a high pressure `coronal' gas outside the disc, its existence has been debated.
Early works on galaxy formation suggested that the collapse of halos should lead to shock heated
gas in the halo. However, various physical processes
during the galactic evolution, such as star formation, galaxy mergers and stripping by intergalactic medium (IGM) would have played a significant role in shaping the extra-planar material that is
observed today. Recent observations have detected this extra-planar gas in HI \citep[e.g.][]{Swaters1997} in H$\alpha$ \citep{Rossa2004,Voigtlander2013} and in X-rays \citep{Wang2001,Strickland2004}. The extra-planar gas, and its different phases, appear in the literature under different names, a few  of which are listed below.

{\it Hot halo gas}: In the standard cold dark matter scenario of structure formation in the universe, baryonic gas falls into dark matter potentials and gets heated to the virial temperature \citep{Silk1977,White1978,White1991}. The gas then cools radiatively, and if the temperature is such that the cooling is rapid (for $T \le 10^6$ K), for the case of low mass galaxies, then no accretion shock develops outside the evolving disc, and most of the halo gas remains at a temperature lower than the virial temperature \citep{Binney1977,Birnboim2003}. This scenario of 'cold accretion' for low mass galaxies has received some observational support in recent years (e.g., Smith et al 2008). In massive galaxies, the hot halo gas is believed to cool slowly, and eventually form warm ($10^4$ K) clouds embedded in a large scale hot corona. These clouds could fall down on the disc in the form of high velocity clouds \citep[e.g.][]{Maller2004,Kauffmann2006}. Numerical simulations have also shown that disc galaxies to be embedded in a hot gaseous halo (Toft et al 2002; Keres et al 2005), and that the X-ray luminosity of the halo gas should scale with galactic mass. However this hypothetical hot halo gas is yet to be observed \citep[e.g.][and references therein]{Rasmussen2009}, and 
the observations of the extraplanar gas so far have been limited to regions close to the disc and bulge, or around active star forming regions \citep[e.g.][]{Wang2007}. 
Moreover, if the halo does contain a large amount of gas, it 
could potentially explain the missing baryon problem, which states that more than $80\%$ of 
the baryons are unaccounted for by collapsed gas and stars in galaxies \citep{Fukugita1998,Anderson2010}. 

{    
{\it Galactic wind}: The subsequent evolution of the halo gas is thought to be governed both by  the inflow of gas from
the IGM and energy injection processes in the disc. Many star forming galaxies have been observed to contain a large amount of outflowing gas \citep[see][for a review]{Veilleux2005}. The early observations of outflowing gas in M82 led to the development of Parker-type steady winds with energy and mass injection from supernovae (SNe) \citep{Burke1968,Johnson1971,Chevalier1985}. These calculations showed that fast steady winds with speeds exceeding $10^3$ km s$^{-1}$ are possible to generate from the
central regions of M82-type starburst galaxies. \citet{Wang1995} explored steady wind solution with radiative cooling and showed that \Osix emission in halos can arise from thermally unstable outflows. \citet{Tomisaka1993} studies the SNe driven outflow and its interaction with halo gas and estimated the extended X-ray emission from M82 type galaxy. Detailed  numerical simulations have  been carried out focused on properties of outflows and their implication for IGM enrichment \citep[e.g.][]{Suchkov1994,Suchkov1996,MacLow1999,Schaye2008,Hopkins2012}. Recently \citet{Mahavir2013} studied steady galactic winds from dark matter halos with Navarro-Frenk-White (NFW) density profile \citep{Navarro1997} and found that SNe can drive outflows from dwarf galaxies, and active galactic nuclei (AGN) power the outflows with speeds exceeding $10^3$ \kms in massive galaxies, e.g., ultra-luminous infrared galaxies (ULIRGs). They also found that winds from intermediate sized galaxies  which are in a quiescent mode of star formation (e.g. the Milky  Way) can not escape the halo. As a consequence of this, one can explain the observed trend of the stellar to halo mass ratio.}

{\it Outflowing cold/warm clouds}: The outflowing gas is often observed to contain a clumpy component that contains neutral or partially ionized atoms, at $10^4$ K \citep[e.g.][]{Martin2005}. In starbursts such as M82, molecular clouds and filaments have also been observed in the outflowing gas \citep{Veilleux2009}. The dynamics of these clouds offer useful clues to the formation and evolution of the outflowing gas, and recently have been used to discuss various physical processes that drives these clouds. \citet{Martin2005} and \citet{Murray2005} suggested that radiation pressure on dust grains embedded in these clouds may play an important role in their dynamics \citep[see also,][]{MahavirFirst,Chattopadhyay2012}. \citet{Mahavir2012} showed that radiation pressure becomes important only for galaxies with mass $\ge 10^{12}$ M$_{\odot}$ and with high SFR, e.g., for ULIRGs. However, the formation process of these clouds remain uncertain. \citet{Marcolini2005} argued that the clouds are likely to be shredded by Kelvin-Helmholtz instability and/or evaporation due to thermal conduction, with a time scale $\le 1$ Myr. It is therefore a puzzle that these clouds are often observed at distances of several kpc, since the travel time (assuming the speed to be a few hundred km s$^{-1}$) is likely to be $\sim 10$ Myr. \citet{Nath2009} and \citet{Murray2011} have suggested a scenario in which radiation pressure pushes a shell of gas and dust which then fragments after being accelerated by thermal pressure from SNe ejecta.

{\it High velocity clouds}: The extraplanar gas is also affected by disc ISM processes in another way.  In the galactic fountain model \citep{Shapiro1976,Bregman1980}, partially ionized gas is launched from the disc by the effect of multiple supernovae (SNe), and after travelling some distance in the halo it falls back to the disc.
An important problem in this regard is that of the formation of clouds in the extraplanar gas, which are observed as either high velocity clouds (HVC) or \Osix clouds.  \citet{Bregman1980} argued that HVCs could form from the condensation of galactic fountain material \citep[but see][]{Ferrara1992}. However, this idea cannot explain the fact that observed metallicity is less than that of the disc ISM \citep{vanWoerden2004}. This is particularly important for HVCs which are more than $\sim 5$ kpc away from the Milky Way disc which have low metallicity ($Z \sim 0.01\hbox{--}0.2 Z_{\odot}$). Recently, \citet{Binney2009} have argued that thermal instability is not efficient in the moving fountain gas, and clouds are  rather likely to form due to the interaction of 
ejected disc material with an pre-existing halo gas \citep[see also][]{Mar2010,Mar2011}. However, the ejected disc material is modeled in a ballistic manner, whereas in reality the outward moving gas is more likely to move like a fluid. 
\cite{Wakker2005} found that \Osix absorption associated with HVCs can arise from the interface between the HVC and
a coronal gas.  
However we note that, in the HVCs nearer than $\sim 10$ kpc, there can be a mixture of components from Galactic ISM, the interface region with halo gas, or gas from the Local group.
Thus the overall picture is tangled because of a contamination at lower heights connected with interaction of falling gas with the extended gas of galactic corona or by the flows generated by internal galactic processes. 

{\it Circumgalactic medium}: The gaseous content of spiral galaxies outside the disc but within the halos of dark matter, formed and shaped by various processes of accretion and outflows, as mentioned above, is generally referred to as the circumgalactic medium (CGM). This refers to 
the reservoir of gas at distances $\sim 100\hbox{--}250$ kpc. However, its existence, and the relation to processes near and in the disc are not yet fully understood. Although outflowing gas is observed near the disc, it is not clear how far this gas propagates, and about the nature of the interaction with pre-existing gas in the halo. It is uncertain  if most of the  wind material escapes the halo, or  is retained in the halo, and then whether or not most of this gas eventually falls back to the disc. A large amount of hot gas in halo can potentially distort the cosmic microwave background radiation through Sunyaev-Zel'dovich effect \citep[e.g.][]{Majumdar2001}. The  low density of the CGM makes it an elusive component to observe, but recently it has been detected through absorption in the line of sight of background  quasars. \citet{Tumlinson2011} detected a large amount of \Osix absorbing gas, at $T\sim 10^{5.5}$ K, in galaxies with moderate SFR. However, its spatial extent and the total gas (and metal) content remains to be measured with enough accuracy that can constrain theoretical models.

In the present paper we address the question of the origin of high velocity clouds, circumgalactic material and the cold clumps in galactic winds, as a result of  the interaction of a steady galactic wind with the relic halo gas distribution using 2D simulations. Our aim is to show that all these constituents  of extra planar material may have a common origin, which is the interaction zone between galactic outflow and hot gas in halo. 

The remainder of the paper is organised as follows. In \S 2 we describe our simulation set-up, including the initial and boundary conditions. In \S 3 we give analytical estimates for the dynamics of the interaction zone, and  compare the
timescales relevant for the problem. 
In \S 4 we present the results of our model runs and then in \S 5 we discuss our results along with the implications.

\section{Simulation set-up}
{    In this work we use the cylindrical ($R,\phi,z$) coordinate system with cylindrical symmetry around the $z$-axis}, in which the distance ($r$) to any random point can be calculated from the in-plane radius ($R$) and height, ($z$) by using $r=\sqrt{R^2+z^2}$. 
The unit of density in our simulation is $1.67 \times 10^{-24}$ g cm$^{-3}$. 
Unit of velocity is 100 \kms and the unit of distance is kpc. Hence the unit of time in the present simulation is 9.5 Myr.
\subsection{Initial and boundary conditions}
\begin{figure}
\includegraphics[scale=0.4]{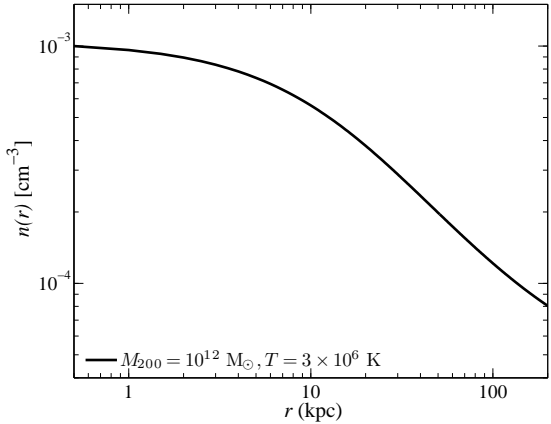}
\caption{The density profile for the gas in the halo.}
\label{fig_denprof}
\end{figure}

We populate the halo with an isothermal gas distribution which is  in hydrostatic equilibrium in a NFW dark matter halo. The gravitational potential for a NFW dark matter halo is given by, \begin{equation}
\Phi(r) = - 2 v_s^2 {\ln(1+ r/r_s) \over  r/r_s}
\label{NFW_pot}
\end{equation}
where $v_s^2 = G M_{200}/[2 r_s f(c)]$ with $f(c)=\ln(1+c)-c/(1+c)$ and $r_s = r_{200}/c$.  Here  $c(=10)$ is the halo concentration parameter, $r_{200}$ is the virial radius and $M_{200}$ is the virial mass. {    We consider a halo of mass $10^{12}$ M$_\odot$ in our simulation.
The corresponding circular speed is $\approx180$ \kms according to the definition given in \citet{Navarro1997}, and the escape speed at the centre is $2v_s$, whose value is $530$ \kms.}  For the gravitational potential in Eq. \ref{NFW_pot}, the hydrostatic density profile is given by 
\begin{equation}
n(r) = n_0 \exp\left[-\frac{\mu m_p (\Phi(r)-\Phi(r_b))}{k T}\right]
\label{eq_haloprof}
\end{equation}
where $n_0$ is the  density at the base, which is also the maximum value of initial density and $\mu (=0.6)$ is the mean molecular weight. We take the density at the base $n_0 = 10^{-3}$ cm$^{-3}$. We note that the similar cored density profiles, with a central density $\sim 10^{-3}$ cm$^{-3}$ have been deduced for the gas in the halo of the Galaxy in recent studies \citep[e.g.][]{Fang2013,Putman2012,Maller2004}. The above profile is characterised by the temperature $T$. We consider a temperature, $T\approx 3\times10^6$ K, which is approximately the virial temperature for a milky way size halo with mass $10^{12}$ M$_{\odot}$ considered in this simulation. Observations also indicate similar temperature for the hot gas in the halo of the Galaxy \citep{Hagihara2010,Fang2013}. 
In Fig. \ref{fig_denprof}, we have shown the density profile of the halo gas for  the halo of mass $10^{12}$ ${\rm M}_\odot$, considered in this work.
{    We set-up a simulation box in which $R$ ranges from $0$ to $R_{\rm max}$ and $z$ goes from $r_b$ to $z_{\rm max}$, where $r_b$ is the launching height. Therefore the simulation box covers  right half of the upper hemisphere of galactic halo. Spiral disc lies at the base of the hemisphere and it lies outside ($z<r_b$) of our computation box .  We use the density profile given by equation (\ref{eq_haloprof}) for outer boundaries at  $z_{\rm max}^+$ and $R_{\rm max}^+$, so that the initial hydrostatic distribution of gas does not change with time. At the lower $z$ boundary, we once again use the density profile given in equation (\ref{eq_haloprof}), except for a region satisfying, $0<\left|R\right|<r_b$, $z< r_b$ in which we inject the material with a specific density and velocity as we discuss below. }

\subsection{Injection parameters}
The injection of wind gas is implemented by assuming that a supersonic wind enters the computation zone from below. We inject (continuously in time) the gas with speed $v_{\rm inj}$ and density $n_{\rm inj}$ in a section of the lower boundary which satisfies $0<\left|R\right|<r_b$. The injected gas has a  temperature corresponding to the sound speed ($c_s$) somewhat lower than  the injection speed ($v_{\rm inj}$) to ensure that the flow is in supersonic regime. This value of $v_{\rm inj}$ will give a wind speed $v_{\rm wind} \sim 2 v_{\rm inj}$, which can be understood simply by the fact that in the supersonic regime, as the wind diverges, the sound speed decreases to small values and the wind speed becomes the sum of initial injection speed and the sound speed at the base.
A supersonic injection right from the base implies that we have assumed a thermalization zone with dimensions $0<\left|R\right|<r_b$, $0<\left|z\right|<r_b$ centred at the galaxy in which the energy and mass injection due to stellar processes  occurs. 
The size of the thermalization zone  ($r_b$) in principle is a free parameter. \citet{Melioli2004} consider $r_b$ in the range $100\hbox{--}700$ pc. In the present study we work with a fixed value of $r_b=500$ pc. 
We use a range of injection speeds and densities. 
{
The injections speed and density are functions of the energy injection rate ($\dot E$), mass loading rate ($\dot M$) and the SFR.
If we consider the thermalized injection zone to be approximately spherical with radius $r_b$, then the mass loading rate for the wind can be approximated as, 
$
\dot M = 4\pi {\rm m_p} n_{\rm inj} v_{\rm inj} r_b^2 
$. Furthmore, one compares this mass loass rate with SFR, by defining a load factor $\eta$ which can be written as,
\begin{equation}
\eta = \Bigl( \frac{\dot M}{\rm M _\odot\, yr^{-1}} \Bigr) \,{\rm SFR}^{-1} 
\label{eq_ett}
\end{equation}
where the ${\rm SFR}$ is in the units of ${\rm M_\odot\, yr^{-1}}$. Energy injection due to the SNe depends on the rate of occurrence of SNe and therefore on the SFR. One can define the energy injection as,  ($\dot E = \epsa f_{\rm sn} E_{\rm sn} {\rm SFR}$) erg yr$^{-1}$. Here $\epsa$ is the efficiency of the energy injection whose value is $0.1$ in normal situations when most of the energy of SNe is lost via radiation and can be as high as $0.5$, for staburst galaxies like M82.  $f_{\rm sn}$ is the fraction of SNe per unit solar mass of star formation, and its value is $1.26 \times 10^{-2}$ for a Kroupa-Chabrier initial mass function \citep{Chab2003}. $E_{\rm sn}=10^{51}$ erg is the energy yield of a single supernova. Using these values we get the energy injection rate, $\dot E = \epsa (4\times 10^{41})\, {\rm SFR}$ erg s$^{-1}$. For the SNe driven wind the energy injection rate is the mechanical luminosity of the wind and it is related to the mass loss rate by  $\dot E = \frac{1}{2}\dot M v_{\rm wind}^2 = 2 \dot M v_{\rm inj}^2$, where we have used, $v_{\rm wind}=2 v_{\rm inj}$. Equating the above two definitions of $\dot E$, we obtain  following  relation  for the efficiency of energy injection ($\epsa$),
\begin{equation}
\epsa =  0.8\, \Bigl( \frac{\dot M}{\rm M _\odot\, yr^{-1}} \Bigr) \Bigl( \frac{v_{\rm inj}}{500\, {\rm km\, s^{-1}}}\Bigr)^2 {\rm SFR}^{-1}
\label{eq_eff}
\end{equation}   
In the 4th column of Table 1, we have provided  the quantity $\eta\times{\rm SFR}$, which is essentially the mass loss rate and in the 5th column we have provided $\epsa \times {\rm SFR}$ corresponding to each run. One can infer the value of $\eta$ or $\epsa$ for a particular case by dividing the values mentioned in 4th and 5th column by the SFR. For example if the SFR$=3$ M$_\odot$ yr$^{-1}$ in case of {\it MG1} we would have $\eta \approx 0.7$ and $\epsa\approx0.2$.

}

\subsection{Details of our runs}
We use the publicly available hydrodynamic code {\sc Pluto}  \citep{Mignone2007}. For our runs we use the solver based on total variation diminishing Lax-Friedrich (TVDLF) scheme supplied with the code. 
 {\sc Pluto} has a well tested implementation of radiative cooling \citep{Tesileanu2008}, which is done by solving the energy equation with a energy loss term, which for this study is a function of density and temperature as given by \citet{Sutherland1993} in case of equilibrium cooling.  We assume solar metallicity for the wind and the halo gas in our simulation. The halo likely has a lower metallicity, but in this work we do not explore the effect of a two-component gas with different metallicities.

{    
We have run the simulation for various values of initial and injection parameters. The quantitative details of parameters involved in our runs are provided in Table 1.  
Various set of parameters are chosen to study the dependence of formation of multiphase medium on hot halo gas and wind properties.  We set a resolution of $250\times500$, $500\times1000$ and $1000\times2000$, for the box of dimensions $10\times20, 20\times 40, 50\times 100$ kpc respectively. Therefore the smallest cell size in our simulations is 40 pc for the box with heights 20 and 40 kpc and a cell size of 50 pc for the box with height 100 kpc. We find that cloud formed in our simulation have size roughly in the range $100\hbox{--} 500$ pc and hence the resolution is adequate to capture the clouds. We have also checked with more finer resolution (cell size=20 pc), and the cloud sizes are unaffected.}
 \begin{table}
 \centering
 \begin{minipage}{140mm}
  \caption{The parameters of our runs}
  \begin{tabular}{@{}cccccc@{}}
  \hline
   Name  &\, \, $\frac{n_{\rm inj}}{{\scriptsize  cm^{-3}}}$ & ${v_{\rm inj} \over {\scriptsize \rm km\, s^{-1}}}$ & $\eta\times{\rm SFR}$  &  $\epsa\times {\rm SFR}$ & ${{\rm box\, size} \over {\scriptsize \rm kpc\times kpc}}$  \\
   
 \hline
 {\it MG1}   & 0.1 & 300 & 2.23 & 0.64& $10\times20$     \\
 {\it MG2}   & 0.1 & 300  & 2.23 &0.64 &  $10\times20$   \\
 {\it MG3}   & 1.0  & 400  & 29.78  & 15.12 & $20\times40$ \\
 {\it MG4}   & 2.0  & 600  & 89.34 & 102.56 & $50\times100$  \\
 {\it MG5}   & 0.5 & 500 & 18.61 & 14.84  & $50\times100$ \\
 {\it MG6}   & 1.2 & 500  & 44.67 & 35.61 & $50\times100$   \\
 {\it MG7}   & 0.05 & 500 & 1.86 & 1.48 & $20\times40$ \\
 {\it MG8}   & 0.4 & 350 & 10.41 & 4.06 & $20\times40 $ \\
 {\it MG9}   & 2.0 & 350  & 52.03 & 20.32 & $20\times40 $ \\
 {\it MG10}   & 0.1 & 400  & 2.97 & 1.52 &  $20\times40$ \\
 {\it MG11}   & 0.1 & 500  & 3.72  & 2.96 & $50\times100$\\
 {\it MG12}   & 0.5 & 500  & 18.58  & 14.81 & $50\times100$ \\
 {\it MG13}   & 0.01 & 200  & 0.15  &  0.02  & $10\times20$   \\
 {\it MG14}   & 10.0 & 300  & 223 & 67 & $10\times20$  \\
 {\it MG15}   & 5.0 & 200  &  74.3 & 9.5 & $10\times20$\\
 \hline

\end{tabular}
\end{minipage}
\caption{List and details of runs :  Combinations of $n_{\rm inj}$ and $v_{\rm inj}$ adopted for our runs are given in 2nd and 3rd column. In the 4th column we have provided  the quantity $\eta\times{\rm SFR}$, which is essentially the mass loss rate and in 5th column we have provided $\epsa \times {\rm SFR}$ corresponding to each run. One can infer the value of $\eta$ or $\epsa$ for a particular run by dividing the values mentioned in 4th and 5th column by the SFR (i.e. for an SFR of ${\rm 3\, M_\odot\, yr^{-1}}$  in case of {\it MG1} we would have $\eta \approx 0.7$ and $\epsa\approx0.2$).  In the last column we have provided the dimensions of simulation box. All of our runs except {\it MG1} have an implementation of equilibrium radiative cooling.  }
\end{table}

\section{Analytic considerations}
In this section we provide analytical estimates relevant for the simulations in this work. This problem has similarities with the case of a wind blown bubble where the stellar wind is expanding in a uniform density medium \citep{Weaver1977}. However there are some subtle differences between the stellar wind blown bubble and the present case of galactic wind interacting with the halo gas. While in the stellar wind blown bubble the ambient medium is denser than the wind, here it is  the opposite, at least for the initial evolution when the wind density is larger compared to the ambient density. Also, the temperature of the ambient medium in this case is $\sim 10^6$ K, which is two orders of magnitude larger than the ISM temperature ($\sim 10^4$ K), relevant for a stellar wind blown bubble. Another difference is the stratification in the density of the ambient halo gas. Because of the stratification, the outermost shock does not slow down quickly and the late time evolution should not be given by a self-similar solution as described in \citet{Weaver1977}. One should instead work out the dynamics after taking the density profile into account as in \citet{Kompaneets1960}. Due to this effect, the free expansion phase of the wind lasts longer and if the wind has enough mechanical luminosity so that it crosses the scale height of the halo gas profile, then it may keep expanding freely with the normal wind speed and eventually escape the halo. We quantify these aspects below.

In the general case of a supernova blast or a wind interacting with ambient medium there are three distinct phases of evolution. Initially there is the free expansion phase. 
When the swept up mass becomes comparable to the mass in the wind then the wind is obstructed by the outside mass and the free expansion phase ends,  the wind enters the next phase of evolution in which the motion can be described by a self-similar solution. This phase ends when radiative losses become significant and the system enters into a momentum driven phase. 

Let us do a simple estimate for the distance ($\rfree$) at which the transition from a free wind phase to self similar phase is expected. It can be computed by equating the swept up mass with the injected mass. For this rough estimate, we assume the injection region to be spherical
with a radius $r_b$, since $r_b$ is much smaller than the length scale of winds involved in our simulations. This gives, 
\begin{equation}
\int_{r_b}^{\rfree} (4\pi r^2) m_{\rm p} n(r) dr = \int_{0}^{\tfree} 4\pi m_{\rm p} n_{\rm inj} v_{\rm inj} r_b^2 dt \,,
\end{equation}
where $n(r)$ is the density profile of the gas in the halo as given by equation ($\ref{eq_haloprof}$) and $\tfree$ is the time which marks the end of free expansion phase. 
The left hand side in the above equation is the swept up mass of the halo gas and the right hand side is the total injected mass till the time $\tfree$. The $\rfree$ is related to $\tfree$ via $\rfree = \int_{0}^{\tfree} v_{\rm wind} dt$. As explained in the \S 2.2 that in the supersonic regime wind speed is,  $v_{\rm wind}\sim 2 v_{\rm inj}$, therefore  $\tfree = \rfree/(2 v_{\rm inj})$.
We find that $\rfree$ is roughly constant for $n_{\rm inj} \le 0.1$ cm$^{-3}$, and
lies in the range of $2 \hbox{--}5$ kpc, but it increases to $\sim 30$ kpc for $n_{\rm inj}\sim 1$ cm$^{-3}$. 
%
This implies that for low injection density, the free expansion phase ends well inside the high density core of the halo gas profile, and the shell will decelerate according to the Weaver et al. (1977) solution. However, for the high
injection density, the free expansion phase would continue for a longer time.
\begin{figure}
\includegraphics[scale=0.4]{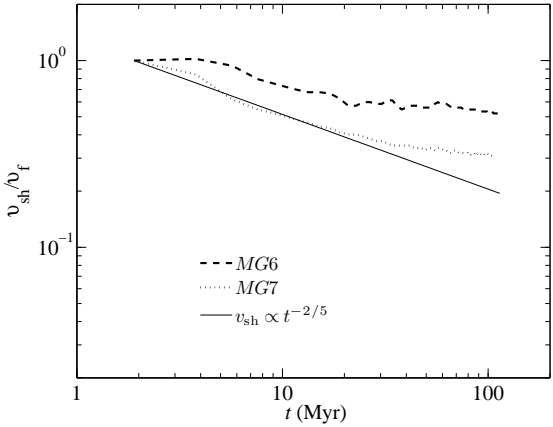}
\caption{Velocity of the forward shock ($v_{\rm sh}$)from our simulation runs {\it MG6} (dashed line) and {\it MG7} (dotted line) is compared with the self-similar solution for a stellar wind blow bubble expanding in a uniform density ISM (solid line). The $v_{\rm sh}$ is shown in the units of velocity of free expansion which is $10^3$ \kms for {\it MG6} and {\it MG7}. 
}
\label{fig_vsh}
\end{figure}

To illustrate this point, we plot in Figure \ref{fig_vsh} the velocities of the forward shocks from two of our simulation runs ({\it MG6, MG7}), with a higher ($n_{\rm inj}=1.2$ cm$^{-3}$) and a low injection density ($n_{\rm inj}=0.05$ cm$^{-3}$) respectively. We show the case of {\it MG6} with a dashed line and that of {\it MG7} with a dotted line, and compare both with the slope of $v_{\rm sh}\propto t^{-2/5}$ (solid line) expected for a stellar wind blown bubble expanding in a uniform density ISM. 
The case of {\it MG6} with large injection density shows less deceleration compared to {\it MG7}. 
If we consider a power-law density profile given by $n\propto r^{-m}$, then in the self-similar phase one would expect $r_{\rm sh}\propto t^{3/(5-m)}$ and $v_{\rm sh} \propto t^{-(2-m)/(5-m)}$. The halo gas profile (Fig \ref{fig_denprof})  beyond the core region can be approximated as a power law decreasing  with $r$, with index $m\sim 1$. One therefore expects $v_{\rm sh}\propto t^{-0.25}$, implying only a mild decrease in velocity with time (or distance), which indeed is seen in Fig \ref{fig_vsh} for the case of {\it MG6} (dashed line). For the case of {\it MG7} (dotted line), with a low injection density, the final velocity is lower because it enters the self-similar phase at the core region of the halo density profile and therefore it exhibits a steeper decrease in velocity initially.

There is one regime in which our results should resemble the self-similar
solution of the `standard stellar wind bubble'. In an extreme situation, for 
very small values of $\rfree$, the wind will decelerate quickly. We can quantify this regime in the following manner. The extent of the free expansion phase is  small for low injection densities. Therefore, the evolution is governed mainly by the self-similar solution, for which the distance of the outermost shock is $r_{\rm sh} = (Lt^3/\rho)^{1/5}$ and the corresponding velocity is $v_{\rm sh} = 0.6(Lt^{-2}/\rho)^{1/5}$, where $\rho = n\ m_{\rm p}$ is the density of the halo gas, whose value is almost constant ($\sim 10^{-3}$ \mpcc) at small distances from the centre. Considering the inverse power law dependence of the velocity ($v_{\rm sh}$) on time, it may so happen that the wind travels a small distance and it decelerates to negligible velocities. In such a case there is no significant large scale wind. Hence for very low mechanical luminosities, when $n_{\rm inj}$ and $v_{\rm inj}$ are both small, we have a small scale wind. Using the mechanical luminosity, $L= (1/2) 4\pi  n_{\rm inj} m_p v_{\rm inj}^3 r_b^2$,  we can write the following general expression for $r_{\rm sh}$ in this case,
\begin{equation}
r_{\rm sh}\approx 7 \, {\rm kpc} \,  \Bigl ({ n_{\rm inj} \over 0.01 \, {\rm\ cm}^{-3} } \Bigr )^{1/2} \, \Bigl ( {v_{\rm inj} \over 250 \, {\rm km/s}} \Bigr )^{3/2} \, \Bigl ({v_{\rm sh} \over 100 \, {\rm km/s}}
\Bigr )^{-3/2} \,.
\label{sssw}
\end{equation}
From the above equation, larger is the value of $n_{\rm inj}$ and $v_{\rm inj}$,  larger will be the value of $r_{\rm sh}$. If we assume that the wind decelerates to a small velocity, say $v_{\rm sh}\approx 100$ \kms, before crossing $r_{\rm sh}\approx 7$ kpc (i.e., the core of the gas density profile in halo), then  the corresponding values of $n_{\rm inj}$ and $v_{\rm inj}$ separates the cases of small scale wind bubble and the large scale galactic wind. Note that the mechanical luminosity corresponding to the limiting case given by the above equation is $L\sim 10^{39}$ erg s$^{-1}$.

We also note that if the $n_{\rm inj}$ is too large such that the total energy lost per unit time due to radiative  cooling ($\dot E_{\rm cool}$) at the base is more than the energy injection rate at the base ($\dot E_{\rm kin} \sim (1/2) m_p n_{\rm inj} v_{\rm inj}^3 4 \pi r_b^2$), then it will lead to catastrophic cooling. To quantify this we can define the following ratio,
\begin{equation}
{\dot E_{\rm cool} \over \dot E_{\rm kin}} \sim {{({4 \pi \over 3} r_{\rm b}^3) (n_{\rm inj} })^2 \Lambda \over m_p n_{\rm inj} v_{\rm inj}^3 2 \pi r_b^2} \sim \Bigl ({ n_{\rm inj} \over 2 \, {\rm  cm}^{-3} } \Bigr ) \, \Bigl ( {v_{\rm inj} \over 300 \, {\rm km} \, {\rm s}^{-1}} \Bigr )^{-3} \,,
\label{eq_catcool}
\end{equation}
assuming $\Lambda \sim 10^{-22}$ erg cm$^3$ s$^{-1}$. We note that the above condition implies a critical mechanical luminosity of $L\sim 10^{42}(n_{\rm inj}/ 2\, {\rm cm^{-3}})(v_{\rm inj}/ {\rm 300\, km\, s^{-1}})^3$ erg s$^{-1}$ above which the cooling should dominate, and in this case the outflow would be strictly momentum driven.

\section{The simulation results}
\begin{figure}
\includegraphics[width=\linewidth]{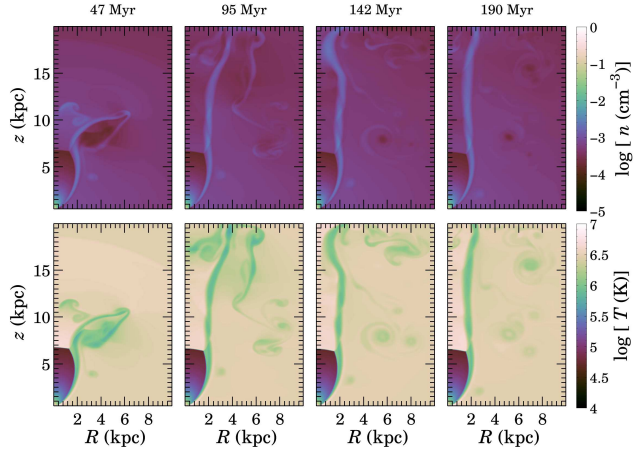}
\caption{Time evolution of log of density (upper panel) and log of temperature (lower panel) in our first run,  {\it MG1}, without radiative cooling. Wind injection velocity is 300 \kms.  }
\label{fig_MG1}
\end{figure}

Here we present the results from five of our runs. To start with we show the case with no radiative cooling in Figure \ref{fig_MG1}, which clearly demonstrates the effect of the halo gas on the wind. One can see the formation of eddies  at the periphery of the wind cone due to the relative motion between the wind gas and the halo gas, and the onset of Kelvin-Helmholtz instability. This leads to turbulent mixing of gas in the halo.  Note that the forward shock in this run reaches a distance of $\sim 20$ kpc in the vertical direction by 47 Myr , and shock heats the ambient halo gas. At this time, there is a contact discontinuity at a vertical distance of $\sim 7$ kpc, where one finds a slightly higher density. 
 The region vertically below this contact discontinuity is occupied by shocked wind gas, and heated by the reverse shock.

If a wind is launched  with a radial injection  in vacuum, it diverges geometrically as it reaches large distances. In other words if the injection and the external forces are spherically symmetric then the wind will keep following the radial streamlines. 
{    In our case although the wind is launched only with $z-$component of velocity, the high pressure of the fluid ensures that the wind diverges as soon as it enters the box.
}

In this first case of ours we can see in Fig. \ref{fig_MG1} that the wind cone does not diverge as much as it should have done if it were to expand freely in vacuum. {    It is halted at a distance of $\sim 7$ kpc because of the resistance offered by the halo gas. The wind is not perfectly conical, and the cross-sectional area becomes fixed roughly beyond $\sim 3$ kpc. Previous studies have shown that the wind may or may not diverge with increasing distance, depending on how stratified the ambient medium is. It has been shown that if the ambient medium has an exponential or Gaussian density distribution then also the wind diverges as it does in vacuum \citep{maclow1989}. However if the decrease in density is not steep, then one would expect a reduction in the cone width at large distances.}

\begin{figure}
\includegraphics[width=\linewidth]{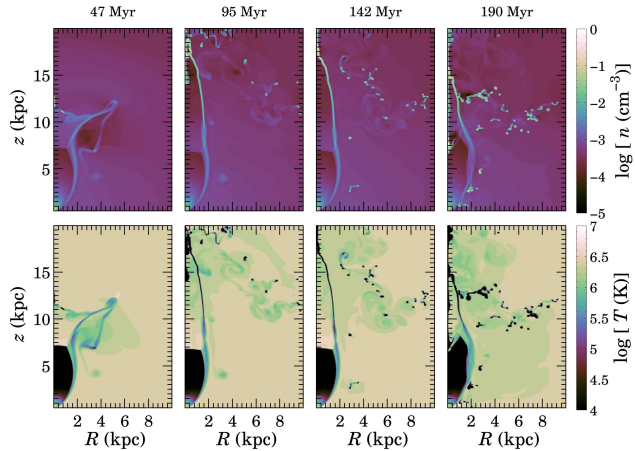}
\caption{Time evolution of log of density (upper panel) and log of temperature (lower panel) in the run {\it MG2}. Wind injection velocity is 300 \kms and injection density at the base is 0.1 cm$^{-3}$. }
\label{fig_MG2}
\end{figure}

In Figure \ref{fig_MG2}, we show the density and temperature evolution for the case {\it MG2}, for which the injection density and velocity are the same as in previous case however radiative cooling is switched on. An additional fact which is now evident is the formation of cold fragments in the high density zones, which are the interaction zones between the halo and wind gas on the periphery of the wind cone. In this particular case ($n_{\rm inj}=0.1$ cm$^{-3}$, $v_{\rm inj}=300$ km s$^{-1}$), the cold clouds that form on the periphery of the eddies remain in the simulation box, but a few clouds that form on top of the wind cone escapes the simulation box by the end of the run. These clouds do not form only from the gas ejected from the disc, but mostly from the mixed gas in the interaction zone between the wind and halo gas. Therefore in the real galactic situations, if the clouds form via this mechanism then they are likely to have a range of metallicities,  
 similar to the case of high velocity clouds in the Milky Way. This therefore represents a variation on the theme of galactic fountain, but with the important difference that clouds form in the interaction between wind and halo gas. 

\begin{figure}
\includegraphics[width=\linewidth]{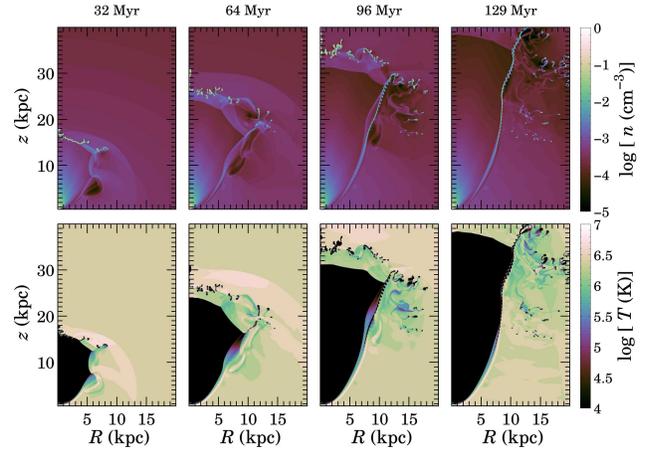}
\caption{Time evolution of log of density (upper panel) and log of temperature (lower panel) in the run {\it MG3}. Wind injection velocity is 400 \kms and injection density at the base is 1.0 cm$^{-3}$. }
\label{fig_MG3}
\end{figure}

Next, we increase the wind density and the size
of the simulation box. In Figure \ref{fig_MG3} we show the case of $n_{\rm inj}=1$ cm$^{-3}$ and wind speed $v_{\rm inj}=400$ \kms. The enhanced wind momentum in this case is able to plough through the halo gas more effectively, as a result of
which we see a diverging wind cone (in contrast to earlier cases). In this case,  we have mostly forward moving gas and even the clumps which form in the interaction zone ride with the flow in the simulation box. The clouds formed in this case represent the case of ram pressure driven clouds in superwinds.
The temperature plots in this run show temperatures less than $10^4$ K in the entire wind cone. This has occurred due to strong cooling in this higher density case. This structure is an outcome of our assumption of single injection site centred at the galaxy. In reality the injection zones may be randomly distributed in the entire disc. In that case the winds  emerging from different injection zones will collide and there will be a more non-uniform density and temperature profile in the region above the disc plane \citep[e.g.][]{Melioli2013}. 

\begin{figure}
\includegraphics[width=\linewidth]{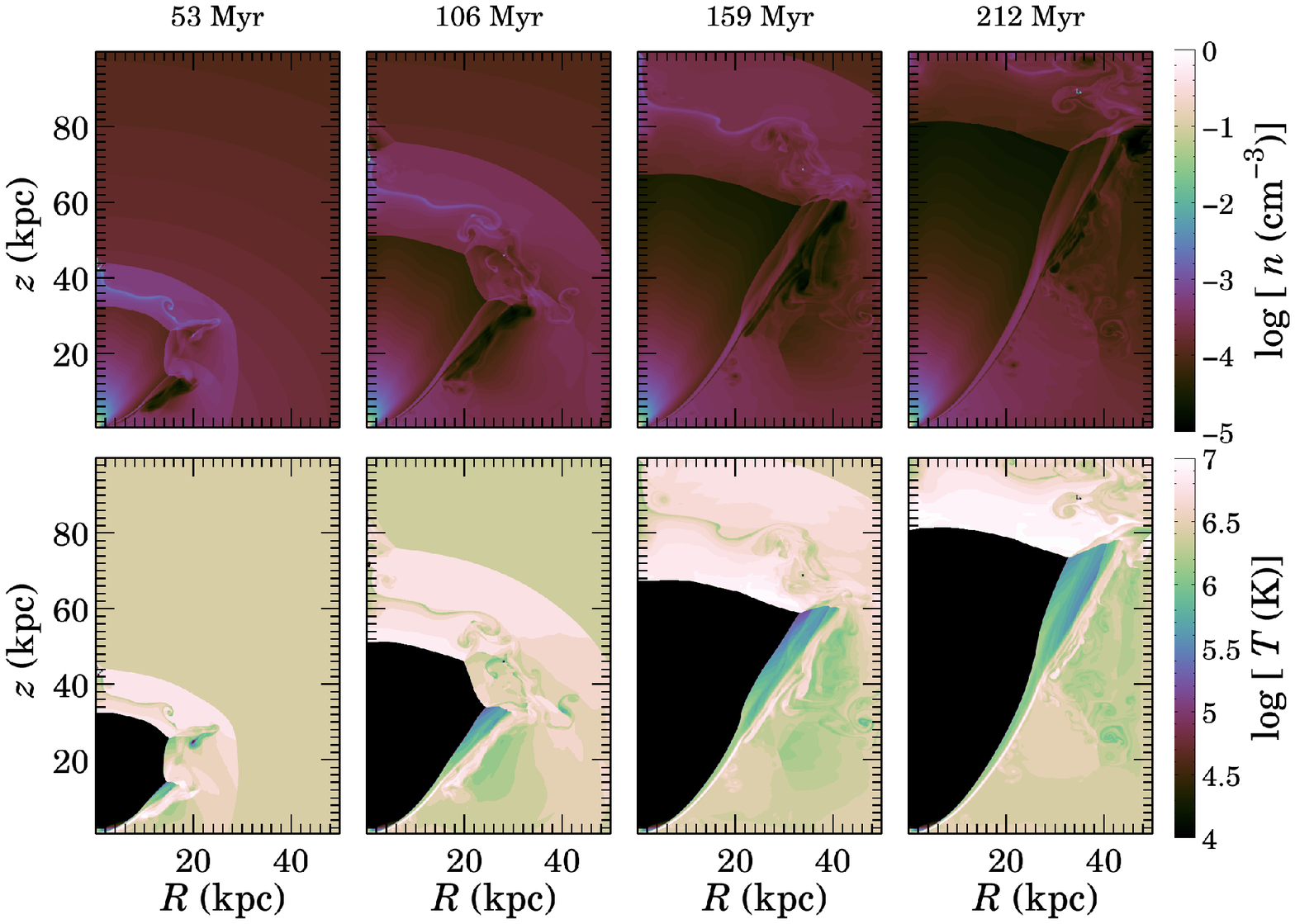}
\caption{Time evolution of log of density(upper panel) and log of temperature (lower panel) in the run {\it MG4}. Wind injection velocity is 600 \kms and injection density at the base is 2 cm$^{-3}$. }
\label{fig_MG4}
\end{figure}

In Figure \ref{fig_MG4}, we increase the wind injection speed to $600$ km s$^{-1}$ and injection density to 2 cm$^{-3}$ and we also increase the size of the simulation box, in order to study whether or not the wind with large injection speed can travel large distances and escape. In this case, the wind clears the simulation box in $\sim 200$ Myr. {    The additional momentum pushes the compressed shell further into the halo, where the shell gas is diluted and does not fragment into clouds.
}

\begin{figure}
\includegraphics[width=\linewidth]{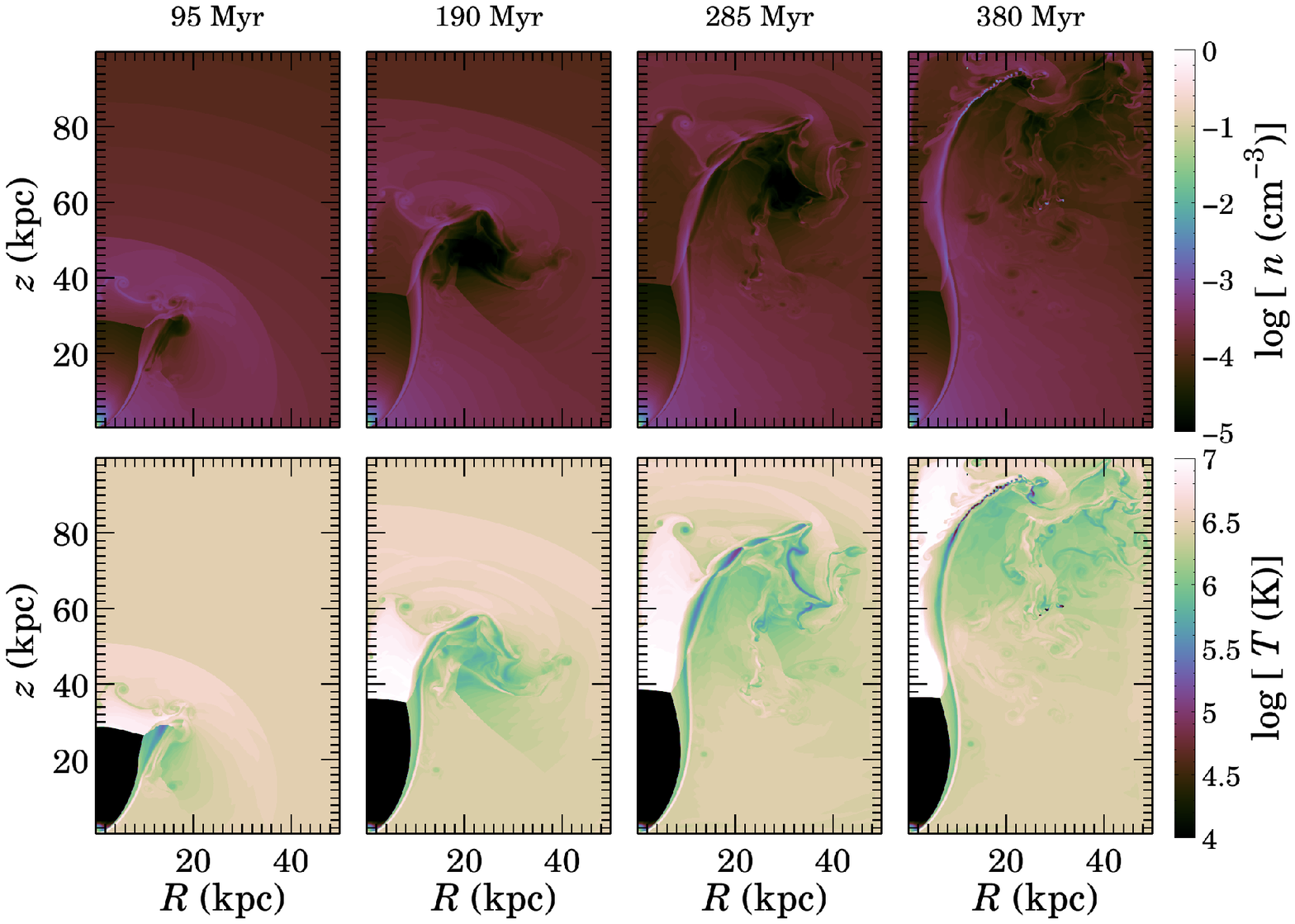}
\caption{Time evolution of log of density(upper panel) and log of temperature (lower panel) in the run {\it MG5}. Wind injection velocity is 500 \kms and injection density at the base is 0.5 cm$^{-3}$. }
\label{fig_MG5}
\end{figure}

Finally, in Fig. \ref{fig_MG5} we decrease the wind density and keep a injection velocity of 500 \kms in a box height of 100 kpc. In this case, radiative cooling is not so effective because of lower density {    as in the previous case}. Also, the wind cannot escape because  at some height the wind density falls below that of the halo gas. As a combined effect of these two aspects, we have regions of low density warm gas ($10^5 \hbox{--}10^6$ K) present in the halo, as shown in the density and temperature snapshots. The column density of this warm gas is $\ge 10^{18}$ cm$^{-2}$, e.g. for a density of $\sim 10^{-4}$ cm$^{-2}$ and a length scale of $\sim 10$ kpc. Hence these regions are likely  candidates for \Osix absorption systems observed repeatedly in the past as a circum-galactic medium in the halos of galaxies \citep{Sembach2003,Savage2003,Tumlinson2011}.

Note that a warm-hot phase ($10^5\hbox{--}10^6$ K) is also present in other cases. For example if we look at the temperature profile of clouds in the lower panel of {\it MG2} and {\it MG3} in Figs. \ref{fig_MG2} and \ref{fig_MG3}, they have a wake of warm gas behind them, which appears greenish and roughly corresponds to a temperature of $10^{5\hbox{--}6}$ K. {    The presence of warm-hot medium in the wakes of high velocity clouds is also suggested recently by \citet{Marasco2013}. Furthermore, observations suggest that HVCs contain different species from ionized to neutral \citep{Lehner2012, Thom2012}. We suggest that different species can be explained considering the variation of temperature around the cloud centre (e.g., in the wake of the cloud).}
Furthermore, the temperature snapshots of {\it MG1} (the simulation run without cooling) show regions with gas at $\sim 10^{5.5}$ K, which implies that if the cooling is suppressed (possibly in systems with low metallicities), one can also generate \Osix absorption systems simply by adiabatic cooling of galactic wind gas that is confined inside the halo.



\begin{figure}
\centering
\includegraphics[width=0.5\linewidth]{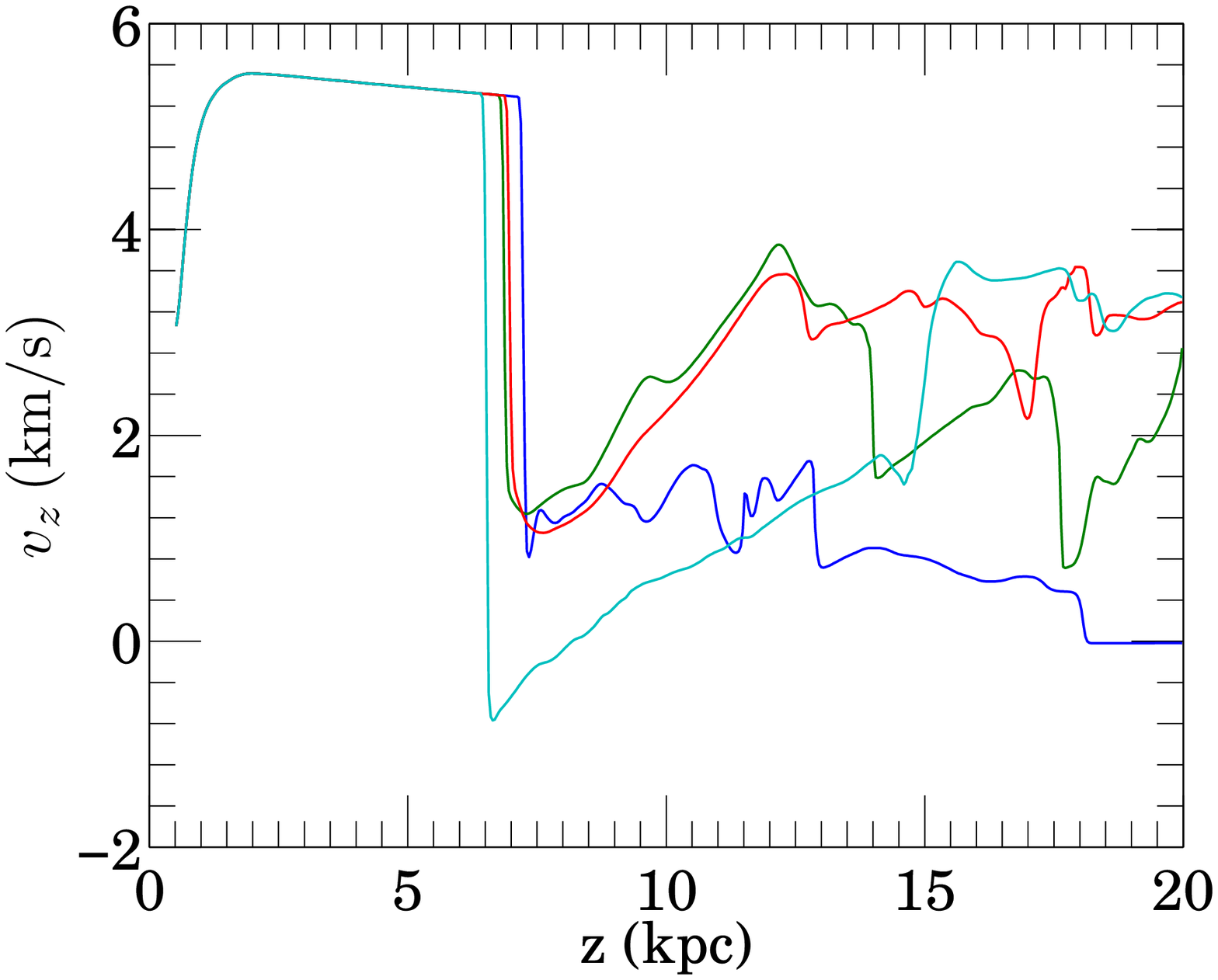}
\includegraphics[width=0.5\linewidth]{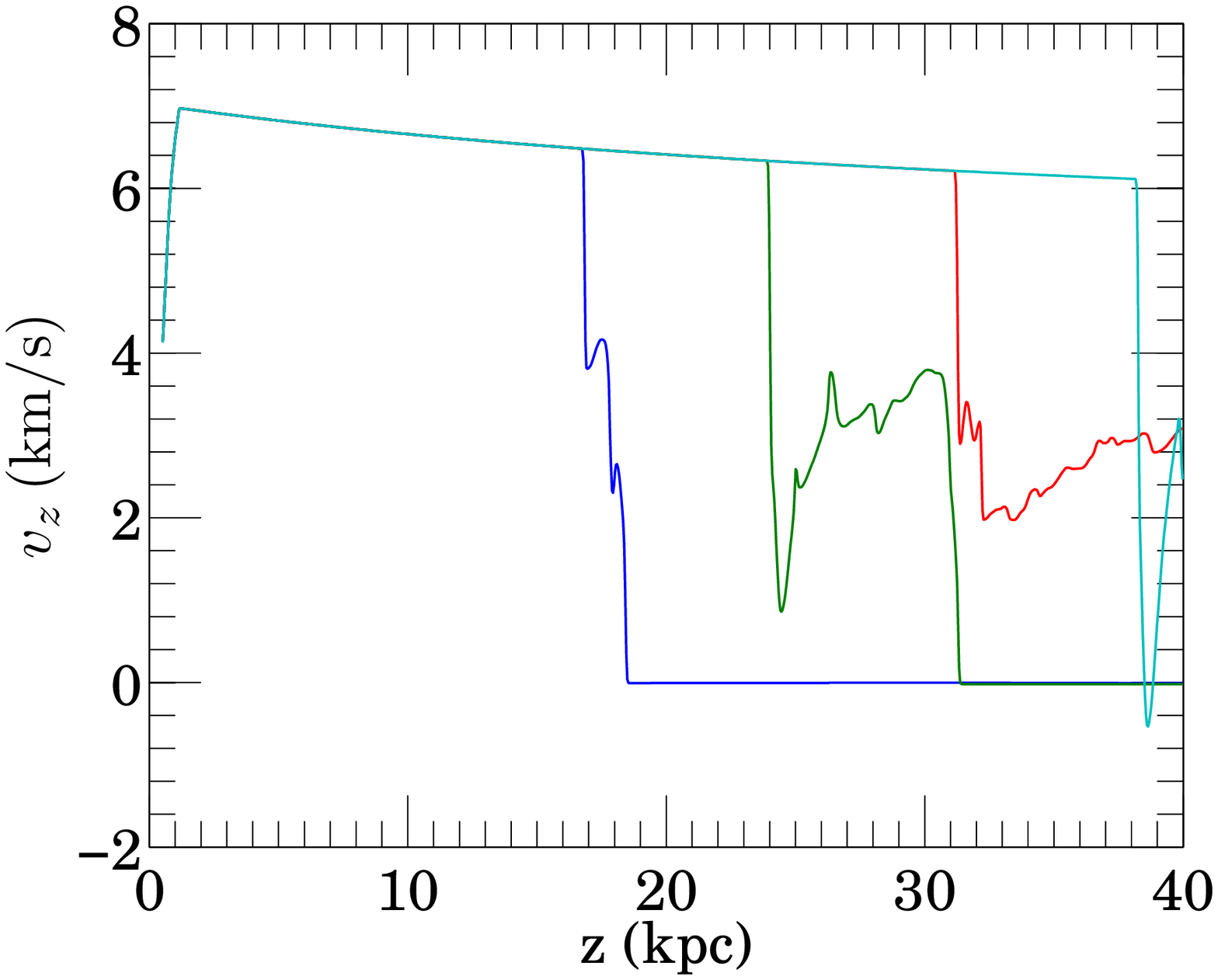}
\\
\includegraphics[width=0.5\linewidth]{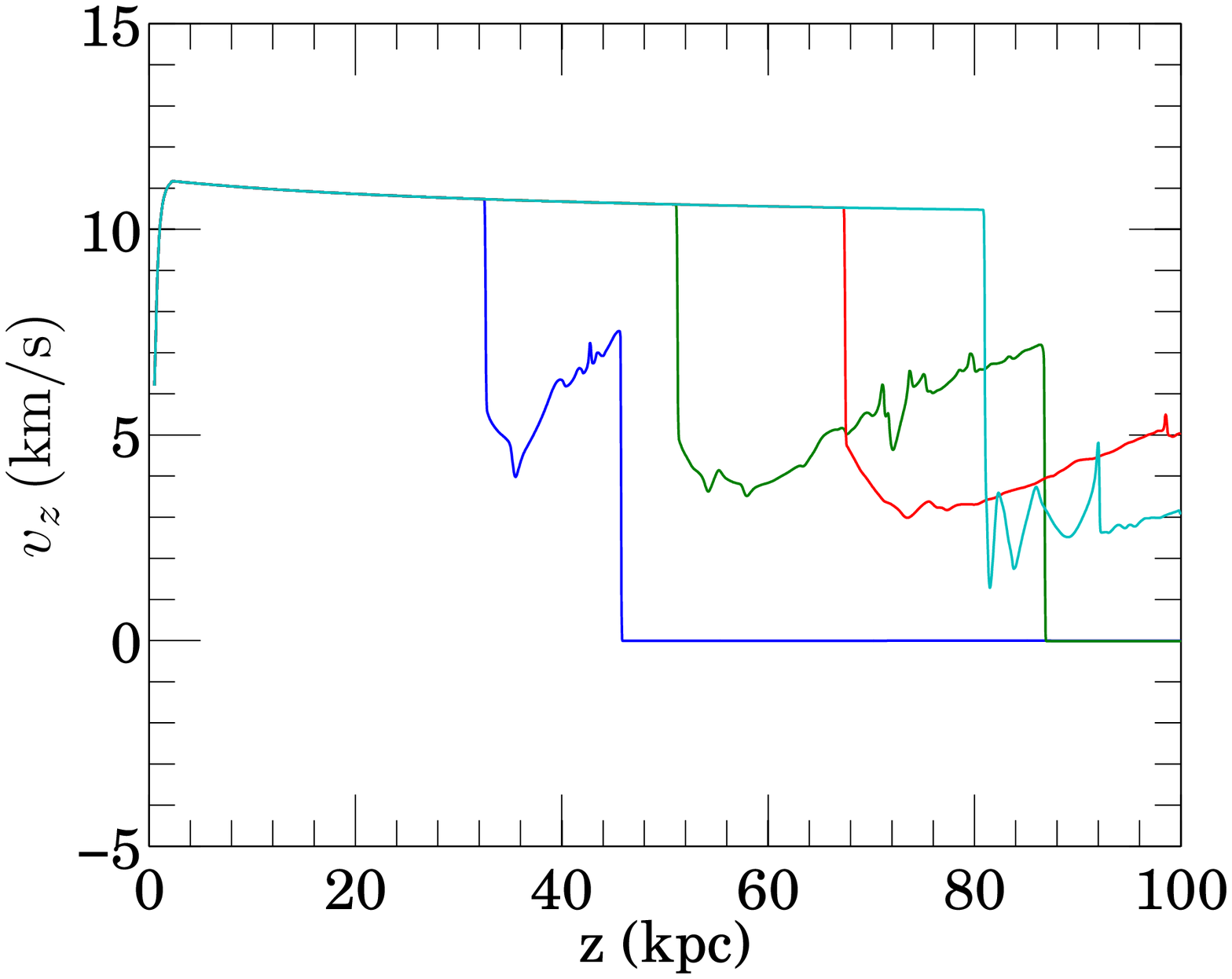}
\includegraphics[width=0.5\linewidth]{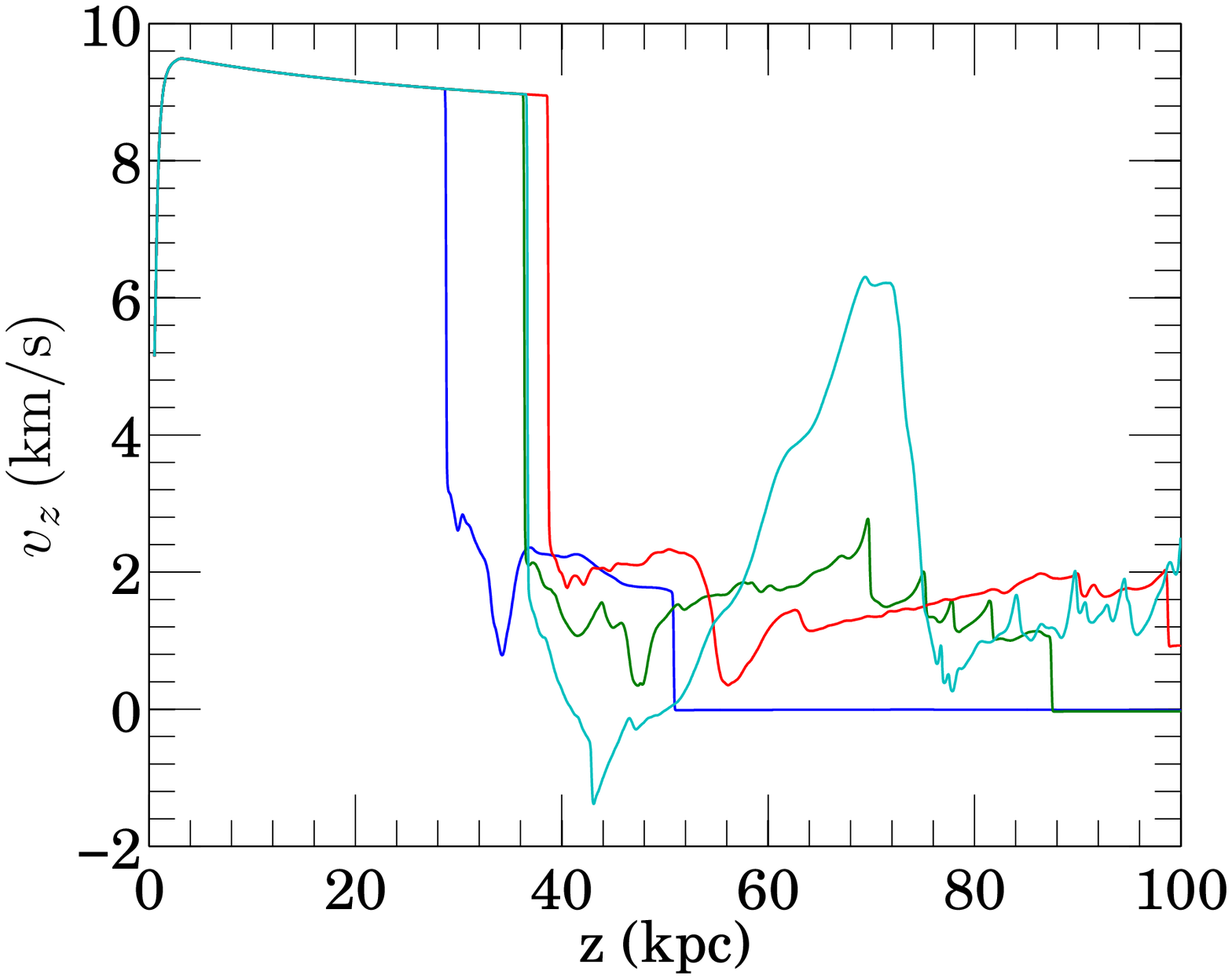}
\caption{Velocity along the z-axis  in computational box at four different times for our  runs {\it MG2} (top-left), {\it MG3} (top-right), {\it MG4} (bottom-left)
and {\it MG5} (bottom-right). For each run four curves are shown with colors blue, green, red and cyan in increasing order of simulation time. The simulation times are the same as used in the density and temperature plots.  }
\label{fig_V}
\end{figure}

We plot the gas velocities as a function of the vertical distance for four snapshots in the case of {\it MG2} (top left panel), {\it MG3} (top right), {\it MG4} (bottom left), {\it MG5} (bottom right)
in Figure \ref{fig_V}. The wind speed initially increases to a value almost double of the injection speed, as expected in a steady (transonic) wind case
\citep{Mahavir2013}, and explained in \S 2.2. The wind speed then drops to a low value at a certain
distance because of the interaction with halo gas. This signifies the boundary between the steady wind and shocked wind that arises due to a reverse shock. 
Then, depending on the efficiency of cooling, there is a formation of thin shell which becomes momentum driven, and 
a consequent rise in the velocity is seen. 
In contrast, for the case of low cooling efficiency (bottom right panel, for the run {\it MG5}), the shocked region is thick and the existence of thin shells 
is not apparent in the velocity plot. 

Another important point to note in the velocity plots is the effect of the halo gas in decelerating the wind front. Compare the bottom-left and bottom-right panels, for the high injection speeds. With a lower injection density in the bottom-right panel, we find that the wind is stuck at  $\sim40$ kpc even
when the injection speed is large enough to overcome the gravitational potential
barrier. This is an example in which the wind is contained within the halo due to the resistance offered by the halo gas.  This aspect is also clear from our discussion in \S 3 where in Fig \ref{fig_vsh} we have plotted the velocities of
the forward shocks for {\it MG6} and {\it MG7} using dashed and dotted lines. Although these two runs have same injection speed, still for {\it MG7}, the velocity of the forward shock becomes almost half of that in {\it MG6}.

We would like to point out that the wind gas cools but does not fragment while the fragmentation occurs only in the dense shell. The reason is that thermal instability grows fast in the compressed high density region while the wind gas diverges on account of its speed, and becomes dilute after entering the simulation box, and the low density does not favour thermal instability.
\subsection{Cloud properties}
\begin{figure}
\includegraphics[height=0.39\linewidth]{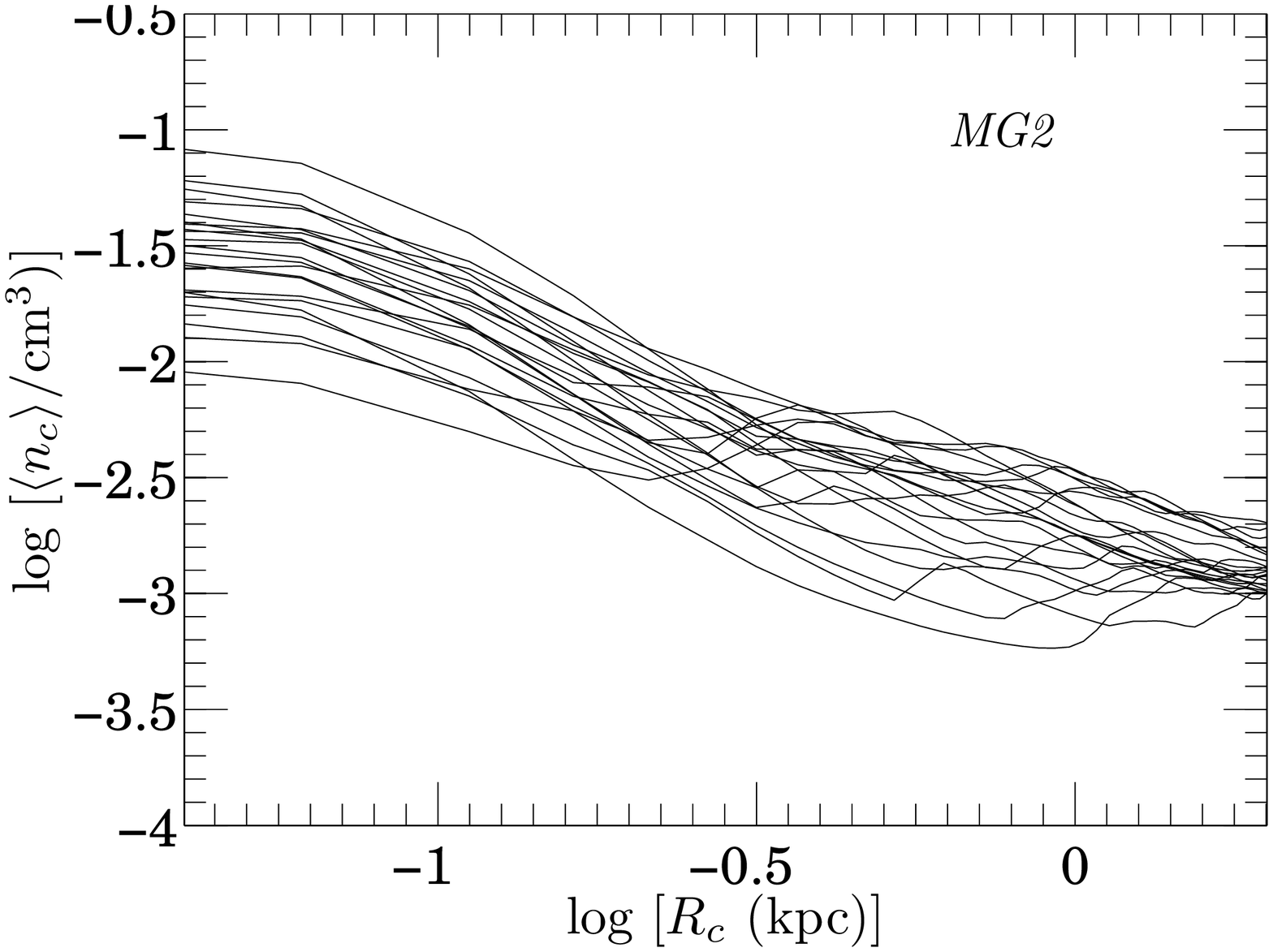}
\includegraphics[height=0.39\linewidth]{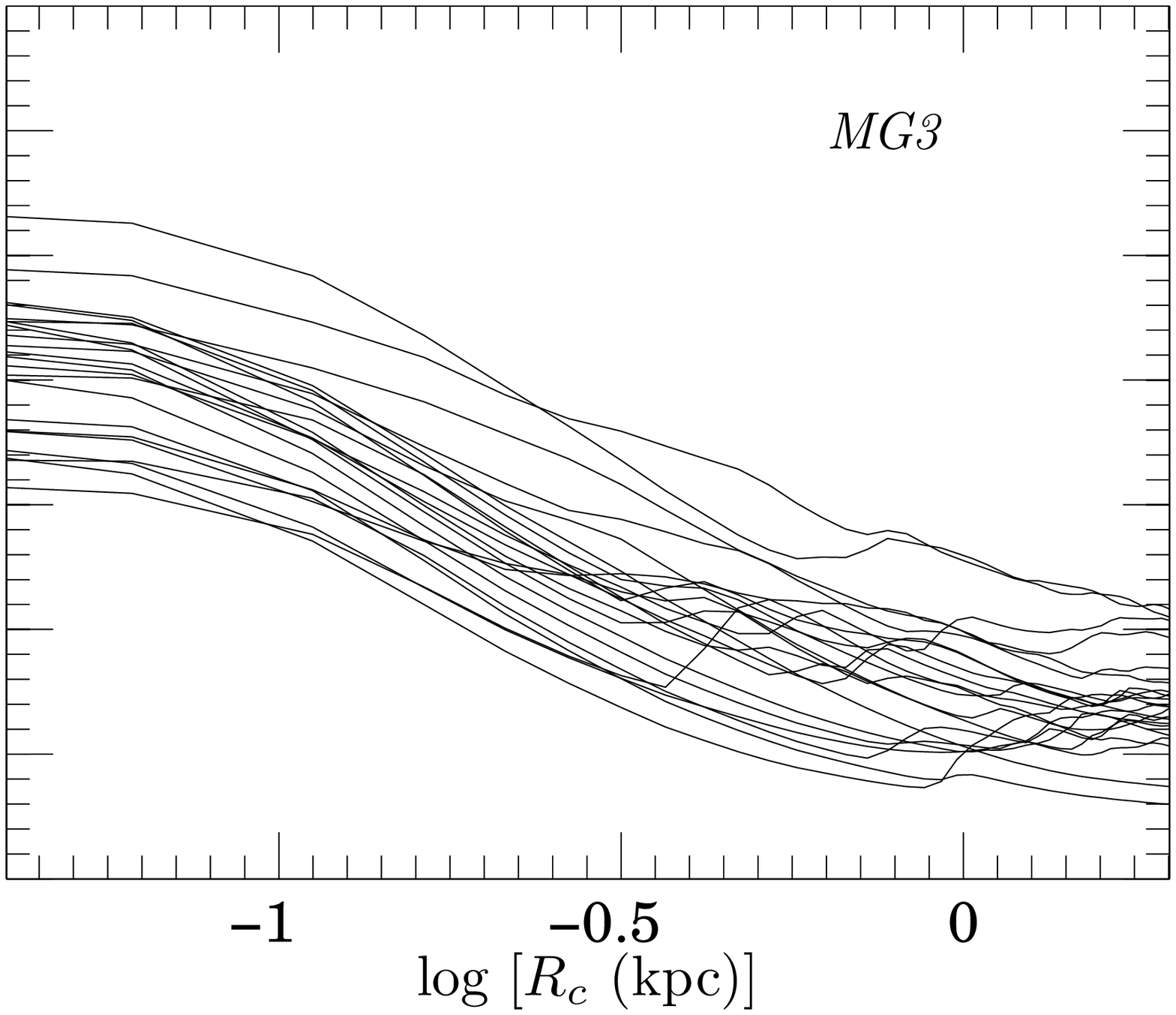}
\caption{Mean density ($\langle\rho_c\rangle$) in a circle centered at the cloud and having a radius $R_c$ is plotted versus $R_c$. Results for the run {\it MG2} and  $\it MG3$ corresponding to a time,  $t=180 $ Myr, are shown in the left and right panels respectively.  }
\label{fig_cloudsize}
\end{figure}

{    Here we would like to determine the properties of the clouds formed in our simulations. In this regard, we estimate the size and mean density of the clouds. Numerically, this is achieved in a few steps, in which we first consider a particular snapshot of the simulation and find all the local maxima in density. Then we define circles of radius ($R_c$) centered at the position of the local maxima. We start with a minimum value of the $R_c$, which is greater than the size of the smallest computational cell in the simulation and calculate the average density in the circular region. We then keep increasing the value of $R_c$ and determine  the average density ($\langle n_c\rangle$) as a function of $R_c$.

 The result is shown in left and right panels of Figure \ref{fig_cloudsize} corresponding to the runs {\it MG2} and {\it MG3} at $t = 180$ Myr, respectively.  Each line in the plot corresponds to a single cloud.  We can see that, for small values of $R_c$, the densities are high and they decrease slowly with $R_c$,  as we sample the inner high density region of the cloud and slowly move outward.
Afterwards, there is a steep fall in density with $R_c$ as we move out of the cloud region. In the end, the density decreases slowly once again when $R_c$ has become large and the mean density is determined by the gas surrounding the cloud. Size of the cloud is in roughly in the range $100-500$ pc as inferred from the steeper part of the curves when the density decrease rapidly, which implies the transition from high density cloud to low density ambient medium. The estimated size of the clouds ($\gtrsim100$) is greater than the smallest cell size in our simulation which is $40$ pc. Also this estimate for cloud size is consistent with the sizes expected from thermal instability, $t_{\rm cool}\times c_s$, since $t_{\rm cool}\sim 3\hbox{--}10$ Myr and $c_s\sim10\hbox{--}30$ \kms.

The average density of the clouds ($\langle n_c\rangle$) is in the range $10^{-2}\hbox{--}10^{-1}$ cm$^{-3}$. Mass of the cloud can be calculated using $M_c=(4\pi/3)\langle n_c\rangle {\rm m_p} R_c^3$, which comes out to be $M_c=10^{3}\hbox{--}10^5$ M$_\odot$. Considering the fact that roughly $\sim10^2$ clouds are present in the simulation box we find that total mass in clouds is $M_{\rm total,clouds}\sim 10^5 \hbox{--}10^{7}$ M$_\odot$. We note that observations also find a similar amount of mass contained in HVCs \citep{Wakker2007}, although in the simulations the total mass in HVCs is model dependent. We would like to mention here that a few of the curves in Figure \ref{fig_cloudsize}, show abrupt upturn for large $R_c$, which is due to the presence of a neighbouring cloud.
}
\begin{figure}
\includegraphics[width=\linewidth]{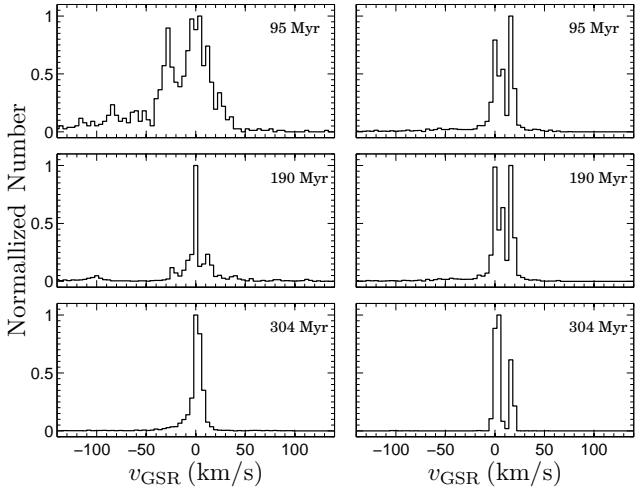}
\caption{Normalized  number of high density zones (clouds) as a function of the velocity in galactic standard of rest frame. Left panel corresponds to our run {\it MG2} at various times of evolution, and the right panel correspond to {\it MG3}. }
\label{fig_hist}
\end{figure}

{        
Next we explore the velocity of the high velocity clumps formed in our simulations. We determine the velocity of regions with density $n> 0.02$ cm$^{-3}$ in the galactic standard of rest (GSR) frame. We plot the histogram of the velocities thus obtained in Fig. \ref{fig_hist}. Three plots in the left column of the figure correspond to {\it MG2}  and in three plots in the right column represent the case of {\it MG3}. We have shown the velocity distribution at three different epochs, $t=95,190$ and $304$ Myr from top to bottom respectively.

 For the case of {\it MG2} (left column of plots) we obtain a normal distribution for the velocities.
 It is interesting to note that observations of HVCs in the Galaxy also indicate a normal distribution of velocities in GSR frame,
  with a mean velocity of $\approx -50$ km s$^{-1}$ \citep{Blitz1999}. Comparing with the distribution found in our simulation in Fig. \ref{fig_hist}, we can speculate that the HVCs in the Milky Way are created within the last $\sim 100$ Myr, since the left-uppermost panel does show a significant contribution to negative velocities. It is
also possible that the halo gas distribution in Milky Way differs from the simple profile assumed here, which
may lead to differences in the dynamics of clouds. However, the overall agreement between our simulation result
and the observed distribution of HVCs is encouraging.
In case of our run {\it MG3} (right column of plots) we find that a significant number of clouds have positive velocity with respect to the galactic centre. The distribution appears to be bimodal with a normal component centred at a mean velocity of roughly zero and another component having positive velocities indicating outflow. This is because of the fact that the mechanical luminosity of injected wind in case of {\it MG3} is high, resulting in a greater number of outflowing clouds.


We conclude that with a suitable combination of $n_{\rm inj}$ and $v_{\rm inj}$, which in turn depend on the SFR, wind loading factor and the efficiency of energy injection from SNe, one can explain the cold clumpy outflows in galaxies with high SFRs as well as the properties of HVCs in quiescent galaxy such as ours.}  

\section{Discussion}
\begin{figure}
\includegraphics[width=\linewidth]{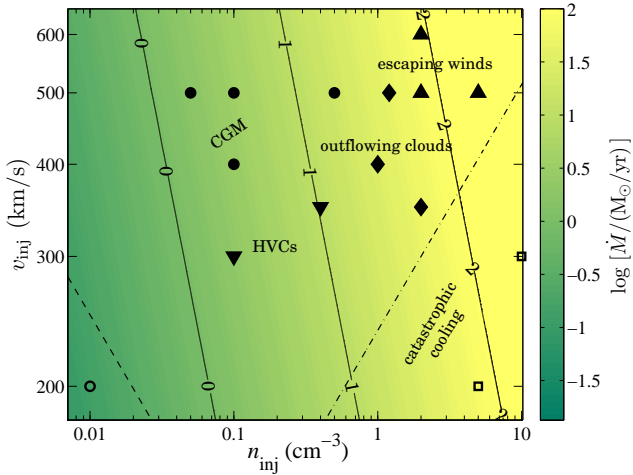}
\caption{ Simulation runs in the parameter space of $n_{\rm inj}$ and $v_{\rm inj}$. The dashed and dash-dotted lines represent the boundaries of region for small scale wind bubble (bottom left corner) and catastrophic cooling (bottom right corner) respectively.  {    Logarithm of wind mass loading rate ($\dot M = \eta\times {\rm SFR}$), is shown by colour coding, with bright yellow regions representing large values of $\dot M$. Three parallel solid lines are the contours corresponding to $\log (\dot M/{\rm M_\odot\, yr^{-1}}) = 0, 1, 2$ from left to right respectively. } } 
\label{last}
\end{figure}

We summarise our results by representing the outcome of different simulation runs in the parameter space of $n_{\rm inj}-v_{\rm inj}$ in Figure \ref{last}. For this purpose we use the results of all our runs mentioned in Table 1, besides the ones discussed in the previous section. Based on our discussions so far, we can first exclude two regions which do not show significant amount of multiphase extraplanar gas. They are: 
\begin{itemize}
\item {\it Small scale wind bubble}: As discussed in the beginning of our analytical estimates (\S 3), if the shell decelerates to negligible speeds within a distance $\sim 10$ kpc, then the wind is confined to small scales and we refer to the case as small scale  wind bubble.
At later times the wind material in this case collapses on itself. We mark this case with an open circle in Figure \ref{last}. {    We also note that for the cases such as these the efficiency of energy injection in the wind and the wind mechanical luminosity would also be unphysically small ($\epsilon\approx 10^{-2}, L_{\rm mech}\lesssim  10^{39}$ erg s$^{-1}$).
The limiting combination of $n_{\rm inj}$
and $v_{\rm inj}$ that leads to such cases can be obtained from equation (\ref{sssw}) by requiring $r_{sh}\le 7$ kpc, and one finds a scaling of $v_{\rm inj} \propto n_{\rm inj}^{-1/3}$ which divides this small scale wind with the galactic wind. We show this limiting combination as a solid line in the bottom-left corner of Figure \ref{last}. 

}

\item {\it Catastrophic cooling}:
If the injection density is very large, then the energy loss by cooling can
be catastrophic and the wind material enters the simulation box and cools quickly before it can diverge into a cone. Two of our simulation runs ({\it MG14} and {\it MG15}) show this 
characteristic and are marked by open squares in Figure \ref{last}.
The corresponding criteria for onset of catastrophic cooling can be inferred from equation (\ref{eq_catcool})  by setting
$\dot{E}_{\rm cool} < \dot{E}_{\rm kin}$. The boundary line for this case has a scaling of $v_{\rm 
inj} \propto n_{\rm inj}^{1/3}$, and is shown in Fig \ref{last} by a dash-dotted line. 
\end{itemize}

Apart from the combination of $n_{\rm inj}$ and $v_{\rm inj}$ for the above two regions, we find significant fragmentation in the interaction zones in other cases.
 In our runs {\it MG2} and {\it MG3}, the interaction zone fragments rapidly into cold clouds of $T\sim 10^4$ K. In {\it MG2}, as the wind gas ploughs through the halo gas, eddies are formed in the side walls
 due to the relative motion. These eddies break out of the bulk flow and form cool clouds. As discussed earlier, we have found the total mass enclosed in these clouds
to be $\sim 10^5\hbox{--}10^7$ M$_{\odot}$ and they extend out to $\sim 10$ kpc from the Galactic centre.  These clouds represents the case of high velocity clouds and this
suggestion is also supported by the comparison of the velocity distributions of HVCs and clouds in runs similar to {\it MG2}.  
We mark these cases with triangles pointing downward in Fig 10 (e.g., {\it MG2}). 

If the injection speed and density are high, as in {\it MG3}, then the wind cone diverges up to a large distance. The clouds formed in these cases (marked with diamonds in Fig 10) are entrained in the bulk flow. Therefore a large number of clouds have a positive velocity outward from the Galactic centre as indicated by the velocity histogram in
the right panel of Fig \ref{fig_hist}. 
We note that these cases resemble  the scenario of 
cold clouds embedded in steady winds \citep{Mahavir2012,Murray2011}.

In our run {\it MG5}, the interaction between the wind and the halo gas leads to significant regions of warm-hot ($10^5\hbox{--}10^6$ K) gas, and in this case cooling
is not effective enough to produce cold clumps.
The column density of the warm-hot gas in this case is sufficiently large to explain the observed \Osix absorption, as discussed earlier. We mark the cases similar to {\it MG5} in Fig 10 with filled circles. We note here that the $10^5\hbox{--}10^6$ K gas is also visible in the wakes behind the moving clouds in our runs {\it MG2} and {\it MG3}, which
can also give rise to \Osix absorption \citep[see also,][]{Marasco2013}.

Finally, if the wind density and speed are further increased, then the wind crosses
$100$ kpc (our simulation box) and possibly escapes the virial radius. The runs that show these features are marked by triangles pointing upward in the top-right corner of Fig. \ref{last}. The shell in this case fragments as well if the density is sufficiently high and clumps can be driven into the intergalactic medium (IGM). These cases are responsible
for the metal enrichment of the IGM \citep{Nath1997,Madau2001,Samui2008}.

{    

The mass loading rate in the wind can be written as $\dot{M}\sim(4\pi r_b^2) m_p n_{\rm inj} v_{\rm inj}^3$. The mass loading rate is related to the SFR ( $\dot{M}=\eta\ {\rm SFR}$) and values of this quantity for our runs is provided in Table 1. Here in Fig \ref{last} we have plotted this quantity using a colour code for increasing $\dot{M}$ from dark green to bright yellow. If we  consider a constant $\eta$ then the dark green regions correspond to low SFR and the bright yellow regions, to high SFR. We can see
that the simulation runs producing CGM fall mostly in the low SFR region. The escaping winds and wind embedded clouds occupy the high SFR zones, and
the infalling cold clouds occupy the intermediate SFR regions. For the Milky Way with a SFR$\sim 3$ M$_\odot$ yr$^{-1}$, and considering a low $v_{\rm wind}=2v_{\rm
inj}$, it should occupy the lower central region of Fig \ref{last}. The result that this region is dominated by infalling clouds in our simulations, agrees
with the observations of HVCs in our Galaxy. We note that from Table 1, for the Milky Way SFR, and for our run {\it MG2}, we obtain $\eta\sim1$, which implies $\dot M\sim {\rm SFR}$. Also the efficiency of energy injection, $\epsa\sim0.2$, which is small and consistent for a quiescent galaxy such as the Milky Way. We further note that, given a SFR one can determine  the value of $\eta$ from Figure \ref{last} by comaring with observed morphology of extraplanar gas. Similarly one can determine $\epsa$ by using the relation $\epsa=0.8\eta(v_{\rm inj}/500\, {\rm km/s})^2$. 
Therefore this figure can be used in conjunction with equation (\ref{eq_ett}) and (\ref{eq_eff}), as a diagnostic to find physical values of wind parameters such as $\eta$ and $\epsa$, by matching with observations. 
}

{       
We have also shown three contours (solid lines) of constant $\dot M$, corresponding to $\dot M = 10^0, 10^1, 10^2$ M$_\odot$ yr$^{-1}$ from left to right respectively. These contours help us to draw important conclusions about the possible values of $\eta$ and SFR.
 For example, HVCs and CGM are likely to appear when the wind mass loading rate is $\sim 1\hbox{--}10$ M$_\odot$ yr$^{-1}$, and the outflowing clouds and escaping winds are possible when the mass loading rate is $\sim 10\hbox{--}100$ M$_\odot$ yr$^{-1}$. This also agrees with the observations where the cold clumpy outflows are observed in galaxies with high star formations rates  \citep[e.g.][]{Martin2005}. 
}
 
{       
 There is yet another interpretation to this figure. If we consider that the SFR is fixed and the mass loading ($\eta$) varies in the plot box, then the left part of the figure represent weak winds with a small  mass loading and the right part a highly loaded wind. The central part of the Figure \ref{last} is the place where wind has a optimum value of mass loading, so as to have physically plausible winds which  can  explain the HVCs and cold clumpy outflows.  }

If we assume that the initial stage of galactic evolution is marked by high SFR, then we expect those galaxies to occupy the top-right corner
of Figure \ref{last}, the region of escaping winds. As the galaxy loses gas via these escaping winds, the SFR decreases, and one enters the regions
of either to the left or bottom-left of Fig \ref{last} depending on the rate of decrement of SFR. This will determine whether or not these galaxies should
have a significant amount of CGM or HVCs (which may sustain the SFR). Therefore the interaction between the wind and the halo gas is
a crucial factor in galactic evolution.

{       
Thermal instability is suppressed in a freely expanding galactic wind \citep{Ferrara1992}. Recent simulations show that clouds also can not condense spontaneously from the hot gas in the halo \citep{Joung2012,Hobbs2013}. However, here we have found  that  clouds can condense in the compressed region formed due to galactic outflow punching through the halo gas. Therefore, this work can be considered a hybrid scenario of the cloud formation from galactic outflow (fountain) and/or the cloud condensation from a stationary or infalling halo gas.}

{       
We would like to point out following limitations of our simulations. We have assumed a uniform metallicity throughout the simulation box, although at large scales inside the halo, the metallicity can vary. This may have lead to an overestimation of cooling in these calculations. However, we also note that these simulations are in 2D for which the effect of cooling is under-estimated. 
Gas metallicity at large distances from the plane is expected to be lower than in the disk, although it is not
clear by how much. For such conditions all the features of the expanding wind bubble would be similar except the transition of the shock front to radiative phase will take longer time. We wish to address these issues in a future paper.
}

\section{Summary}
{       
We have simulated the interaction of a galactic wind with the hot halo gas. We have found that the wind drives a shell of compressed gas which fragments due to instabilities and leads to the formation of HVCs, \Osix regions and outflowing clouds depending on the amount of cooling. We have categorized the formation of various multiphase media in the parameter space of injection velocity and density, which are the input parameters for our simulations. We have found that clouds formed in our simulations have a total mass of $10^5\hbox{--}10^7$ M$_\odot$, and they have a distribution of velocity similar to that of observed HVCs. For higher density and velocity of injection corresponding to vigorously star forming galaxies, a significant number of clouds are outflowing. Furthermore the circumgalactic material with a temperature in range $10^5-10^6$ K is also seen in our simulations with column density appropriate to give rise to the observed \Osix absorption.   These results lead to the conclusion that HVCs, cold outflows and the circumgalactic medium  can form due to the interaction of  galactic wind with halo gas. The injection density and velocity are linked with the SFR, the mass loading factor of outflows and the efficiency of energy injection by SNe. Therefore, the results deduced in these simulations may serve  as a diagnostic to constrain the feedback efficiency of outflows.
}

\bigskip
Acknowledgements:
The authors acknowledge support from an Indo-Russian project (RFBR grant 12-02-92704-IND, DST-India grant INT-RFBR-P121). For our plots in this work, we have made use of `Cubehelix' color scheme described in \citet{Green2011}. We thank P. Sharma and E. Vasiliev for useful discussions. We thank an anonymous referee for constructive comments on the manuscript.

\footnotesize{
\bibliography{ref_7}
}




\end{document}